\documentclass[12pt,tightenlines,
    onecolumn,
    preprintnumbers,
    amsmath,
    amssymb,
    prd,
    showpacs
]{revtex4}
\usepackage{color}
\usepackage{graphicx}%
\usepackage{amsmath}
\usepackage[normalem,normalbf]{ulem}
\usepackage{amssymb}
\usepackage{bm}
\usepackage{enumerate}

\newcommand{\nn}{\nonumber}
\newcommand{\dsp}{\displaystyle}
\begin{document}
\title{Extraordinary vacuum black string solutions}

\author{Hyeong-Chan Kim}
\email{hckim@phya.yonsei.ac.kr}
\affiliation{Department of Physics, Yonsei University, Seoul 120-749,
Republic of Korea.
}%
\author{Jungjai Lee}
\email{jjlee@daejin.ac.kr}
\affiliation{ Department of Physics, Daejin University,
Pocheon, 487-711, Korea.
}%

\date{\today}%
\bigskip
\begin{abstract}
In addition to the boosted static solution there are two other classes of stationary string-like solutions of the vacuum Einstein equation in (4+1)-dimensions. Each class is characterized by three parameters of mass, tension, and momentum flow along the fifth coordinate. We analyze the metric properties of one of the two classes, which was previously assumed to be naked singular, and show that the solution spectrum contains black string and wormhole in addition to the known naked singularity as the momentum flow to mass ratio increases. Interestingly, there does not exist new zero momentum solution in these cases.

\end{abstract}
\pacs{04.70.-s, 04.50.+h, 11.25.Wx, 11.27.+d}
\keywords{black hole, black string}
\maketitle

\section{Introduction}

After the discovery of Schwarzschild black hole solution~\cite{M}
in general relativity, there has been an enormous increase of
interest in black objects such as black hole~\cite{Q,hawking},
black string, and black $p-$brane~\cite{myers2}.
A close connection appears between the black hole and string theory after the discovery that the excited string state can be regarded as an extremal black string solution of general relativity~\cite{horowitz}.
In the presence of the extremal black string, it would be interesting to ask the presence of its non-extremal solution in connection with the excited string state.
However, Schwarzschild black string solution in (4+1) dimensional spacetime is unstable with respect to linear perturbation of metric due to the Gregory-Laflamme instability~\cite{GL,GL2,GM,choptuik,hirayama}.
Therefore, we can not use the Schwarzschild black string as a kinds of ``stable" outcome.
In this sense, it is interesting to study the full geometrical analysis of string-like vacuum solution of Einstein equation.

The static string-like solution in (4+1) dimensions have been studied extensively~\cite{gross,davidson,ohta,lee}. It was shown that the spherically symmetric solution of the vacuum Einstein equation is characterized by two parameters, the mass $M$ and the tension $\tau$ along the direction of the fifth coordinate. The metric of the static spherically symmetric vacuum black string solution in (4+1) dimensions in Ref.~\cite{lee} is
\begin{eqnarray} \label{lee}
ds^2 &=&-\left|\frac{1-K/\rho}{1+K/\rho}\right|^{
    \frac{2(\chi+1/\sqrt{3})}{\sqrt{1+\chi^2}}}
    dt^2+
    \left|\frac{1-K/\rho}{1+K/\rho}\right|^{
    \frac{2(-\chi+1/\sqrt{3})}{\sqrt{1+\chi^2}}}
    dz^2  \\
&+&\left|1-\frac{K^2}{\rho^2}\right|^2
    \left|\frac{1-K/\rho}{1+K/\rho}\right|^{-
    \frac{4}{\sqrt{3}\sqrt{1+\chi^2}}}
    \left(d\rho^2+
    \rho^2 d\Omega_{(2)}^2\right). \nn
\end{eqnarray}
The parameters $K$ and $\chi$ are related to the mass $M$ and tension $\tau (= a M)$ by
\begin{eqnarray} \label{K:}
K &=&\frac{G_5}{2\sqrt{3}}
    (M+\tau)\sqrt{1+\chi^2}, \\
\chi &=& \sqrt{3}\frac{M-\tau}{M+\tau}=\sqrt{3}\frac{1-a}{1+a},\nn
\end{eqnarray}
where $G_5$ is the five dimensional gravitational constant.
The double Wick rotation $t\rightarrow iz$, $z\rightarrow i t$ of Eq.~(\ref{lee}) leads to a new solution with $\chi\rightarrow -\chi$. Only the solution with $\chi=0$ is invariant under the discrete transformation.
Not all physical properties of the solution is yet understood and is
under investigation~\cite{kang}.
Solutions with $\chi=1/\sqrt{3}$ and $\chi=-1/\sqrt{3}$ denote the
Schwarzschild black string and the Kaluza-Klein bubble
solution~\cite{harmark}, respectively. Except for the two cases, the solution is naked singular.
The area of the naked singularity vanishes, which suggests that the solution will be unstable under the metric perturbation.
Since there is no known stable string-like vacuum solution, it raises a question: "What is the final state of the gravitational collapse of a neutral cylindrical object in (4+1) dimensions?".

On the other hand, the charged black string is known to be stable in the extremal limit~\cite{hirayama,KL}.
Therefore, the conserved quantities such as the angular momentum and the electric charge is tend to stabilize the black string system in their extremal limit.
In the case of linear momentum $P$ along the string direction usually is known not to affect to the stability of the string-like solution~\cite{myers}.
However, if there is a solution in which the conserved linear momentum is not directly linked to the boost symmetry (there exists such a solution e.g. in Ref.~\cite{chodos}), the solution may be stable under the perturbation of the metric in some case.
Especially, Chodos and Detweiler classified the stationary string like vacuum solution with spherical symmetry in (4+1)-dimensions into three classes, the usual boosted solution (class I) of the static one~(\ref{lee}), extraordinary one (class II), and wormhole (class III).
The class III wormhole solution is well analyzed~\cite{chodos,clement} and generalized~\cite{Das}, however the class II is simply mentioned to be naked singular.
We study the stationary string-like solution of class II and investigate the properties of the solution. We show that the solution describes a black string in a wide range of parameters contrary to the static case and is entropically stable.

In Sec. II, starting from the usual boosted solution (class I), we re-derive the class II and class III then analyze the class II in detail. The conserved quantities are written in terms of metric parameters and it is shown that the momentum can vanish only when the tension to mass ration is one.
In Sec. III, we observe the Kretschmann invariant and the boost symmetry of class II and show that the mass to tension ratio can be fixed to one with the boost symmetry.
In Sec. IV, we study the causal structure of the solution to show that the solution is a black string for certain parameter range. We investigate their geometric properties in Sec. V and summarize the results in Sec. VI.
At the end of the article, we add two appendices which deals the boost symmetry and the causal structure.

\section{Development of new class of solutions}

In this section, we construct three classes of stationary solutions, characterized by three parameter families, by generalizing the static solution~(\ref{lee}). One of the classes is that of the boosted static solution, however, other two are not directly related to the static one. Even though the original static solution is naked singular, its generalizations contain solutions free from the naked singularity.

A simplest stationary solution can be built by boosting,
\begin{eqnarray} \label{eq:boost:asym}
\left(\begin{tabular}{c}
    ${t'}$\\ ${z'}$ \\\end{tabular}\right)
    &=&\left(\begin{tabular}{c}
    $\cosh \xi~~ \sinh\xi$\\ $\sinh\xi~~
        \cosh\xi$ \\ \end{tabular}\right)
\left(\begin{tabular}{c}
    $t$\\ $z$ \\\end{tabular}\right),
\end{eqnarray}
the static one~(\ref{lee}) with the velocity $v=\tanh \xi$ along the fifth $z-$coordinate.
The boosted metric, dropping the primes after the substitution of the coordinates transformation~(\ref{eq:boost:asym}) to Eq.~(\ref{lee}), is
\begin{eqnarray} \label{metric:ansatz}
ds^2 &=&g_{\mu\nu}dx^\mu dx^\nu=g_{tt} dt^2+ 2 g_{tz} dt dz+
    g_{zz} dz^2
+G(\rho)\left(d\rho^2+\rho^2d\Omega_{(2)}^2\right),\nn
\end{eqnarray}
where the metric components are
\begin{eqnarray} \label{Sol}
g_{tt}&=& D^{-\frac{2}{\sqrt{3}\sqrt{1+\chi^2}}}
    \left(s \,D^{\frac{2\chi}{\sqrt{1+\chi^2}}}-c\, D^{-\frac{2\chi}{\sqrt{1+\chi^2}}}
    \right),\\
g_{zz}&=&
    D^{-\frac{2}{\sqrt{3}\sqrt{1+\chi^2}}}
    \left(c \, D^{\frac{2\chi}{\sqrt{1+\chi^2}}}-
    s \,D^{-\frac{2\chi}{\sqrt{1+\chi^2}}}\right) \nn ,\\
g_{tz} &=&-\sqrt{cs} \,D^{-\frac{2}{\sqrt{3}\sqrt{1+\chi^2}}}
    \left(D^{\frac{2\chi}{\sqrt{1+\chi^2}}}-
     D^{-\frac{2\chi}{\sqrt{1+\chi^2}}}\right),\nn \\
G(\rho)&=&\left(1-\frac{K^2}{\rho^2}\right)^{2}
    D^{\frac{4}{\sqrt{3}
        \sqrt{1+\chi^2}}} . \nn
\end{eqnarray}
For simplicity, we use the notation
\begin{eqnarray} \label{eq:D}
 D(\rho) = \frac{1+K/\rho}{1-K/\rho}
\end{eqnarray}
and the boost is parameterized by the coefficients $c=\cosh^2\xi$ and
$s=\sinh^2\xi$.
The parameters $K$, $\chi$, and $\xi$ are related with the physical
quantities $M$, $\tau$, and $P$~\cite{kim}.

An interesting observation here is that the metric~(\ref{Sol}) is still satisfied with the Einstein equation for complex number coefficients $c$ and $s$ with
\begin{eqnarray} \label{eq:cs}
c-s =1.
\end{eqnarray}
With this observation, we may find two additional sets of real metric solutions by appropriately choosing complex parameters $c$, $s$, $\chi$, and $K$. We call the two by ``class II" and ``class III":
\begin{eqnarray}
\mbox{class I}: c&=& \cosh^2 \xi, ~~ s= \sinh^2\xi, ~~ \xi,\chi, K \in R \,, \\
\mbox{class II}: c&=&\frac{1}{2}- i q,~~s=-\frac{1}{2}-i q,~~\chi = -i \bar
    \chi; ~|\bar\chi|\leq 1,\quad q,\bar \chi, K\in R \,, \nn\\
\mbox{class III}: c&=&\frac{1}{2}- i q,~~s=-\frac{1}{2}-i q,~~\chi = -i
    \bar \chi; ~|\bar \chi|\geq 1,~~K= i Q, \quad q, \bar \chi,Q\in R . \nn
\end{eqnarray}
These solutions were, in fact, noticed by Chodos and Detweiler~\cite{chodos}.
The class III was identified as a regular wormhole solution~\cite{chodos} and is generalized to $n+p$ dimensions~\cite{clement,Das}.

We concentrate on the class II in this paper. Even though this class was briefly mentioned to be naked singular~\cite{chodos}, we find that there is a nontrivial region of parameters which present solutions free from the naked singularity.
After setting
\begin{eqnarray*}
\tan\mu=\frac{\bar \chi}{\sqrt{1-\bar \chi^2}}\,,
\end{eqnarray*}
the metric takes the form,
\begin{eqnarray} \label{Sol3}
g_{tt}&=&-
D^{\frac{-2}{\sqrt{3}\cos\mu}}\left[\cos({2\tan\mu}
    \log D)+ 2q \sin({2\tan\mu}
    \log D)\right] , \\
g_{zz} &=& D^{\frac{-2}{\sqrt{3}\cos\mu}}\left[\cos({2\tan\mu}
    \log D)- 2q \sin({2\tan\mu}
    \log D)\right] \nn ,\\
g_{tz} &=&(1+4 q^2)^{1/2}
    D^{-\frac{2}{\sqrt{3}\cos\mu}} \sin ({2\tan\mu}
    \log D) , \nn\\
G(\rho)&=&\left(1-\frac{K^2}{\rho^2}\right)^{2}
    D^{\frac{4}{\sqrt{3}\cos\mu}} . \nn
\end{eqnarray}
The double Wick rotation $t\rightarrow i z$ and $z\rightarrow i t$ is equivalent to the parameter change $q \rightarrow -q$. The space inversion $z\rightarrow -z$ corresponds to $\mu \rightarrow -\mu$ with $q\rightarrow -q$. The coordinate change $\rho \rightarrow -\rho$ is equivalent to the parameter change $K\rightarrow -K$. The metric is invariant under the change of parameters $K\rightarrow -K$, $\cos\mu \rightarrow -\cos\mu=\cos(\pi-\mu)$, and $q\rightarrow -q$. This symmetry allows one to restrict the range of $K$ being a nonnegative number, therefore, we restrict our attentions to the metric with $\rho \geq 0$.

Asymptotically to $O(K/\rho)$, the metric behaves as
\begin{eqnarray}
g_{tt}&\simeq& -1+\left(\frac{1}{\sqrt{3}\cos\mu}-
    2 q \tan\mu\right)\frac{4K}{\rho},~~
g_{zz} \simeq  1-\left(\frac{1}{\sqrt{3}\sec\mu}+
    2q\tan\mu\right)\frac{4K}{\rho}, \\
g_{tz}&\simeq&
    (1+4q^2)^{1/2}\tan\mu\frac{4K}{\rho} \,, ~~~~~~\quad
G\simeq 1+\frac{8 }{\sqrt{3}\cos\mu}\frac{K}{ \rho}. \nn
\end{eqnarray}
Comparing this with the asymptotic form of metric around a stationary matter source given in Ref.~\cite{kim},
\begin{eqnarray}
g_{tt}&\simeq& -1+\frac{4G_5M(2-a)}{3\rho},~~
g_{zz} \simeq 1+\frac{4G_5M(1-2a)}{3\rho}, \\
 g_{tz}&\simeq& \frac{4G_5 P}{\rho}, ~~ G \simeq 1+\frac{4G_5M(1+a) }{3 \rho}\, ,\nn
\end{eqnarray}
we determine the mass, tension, and momentum flow in terms of the parameters in the metric:
\begin{eqnarray} \label{relation}
M &=& \frac{\sqrt{3}-2q\sin\mu}{\cos\mu} \frac{K}{G_5},\\
\tau &=&\frac{\sqrt{3}+2q\sin\mu}{\cos\mu} \frac{K}{G_5},\nn \\
P &=&\sqrt{1+4q^2}\tan \mu \frac{K}{G_5} \nn .
\end{eqnarray}
Note that for $\mu \neq 0$, the momentum flow $P$ cannot vanish.
The tension to mass ratio, $\displaystyle a= \frac{\sqrt{3}+2q\sin\mu}{\sqrt{3}- 2q \sin\mu }$, depends only on $q\sin\mu$. The $\mu=0$ solution corresponds to the static solution~(\ref{lee}) with $a=1$.
For $q=0$, it also gives $a=1$. However, it does not correspond to the static solution if $\mu \neq 0$.
If we require the positiveness of the mass $M\geq 0$ and the tension $\tau \geq 0$, we have the restriction $|q|\leq \frac{\sqrt{3}}{2|\sin\mu|}$ for $K\cos \mu \geq 0$. On the other hand, we must have $|q|> \frac{\sqrt{3}}{2|\sin\mu|}$ for $K\cos \mu < 0$.

Inverting Eq.~(\ref{relation}), we obtain
\begin{eqnarray} \label{eq:relation}
K &=& G_5M
    \sqrt{\frac{a^2-a+1}3-\left(\frac{P}{M}\right)^2} , ~~~\\
q &=& \mp \frac{1-a}{4\sqrt{\left(\frac{P}{M}\right)^2-\frac{(1-a)^2}{4}   }}, \nn \\
\sin\mu &=& \pm \frac{2\sqrt{3}}{1+a} \sqrt{
    \left(\frac{P}{M}\right)^2-\frac{(1-a)^2}4} . \nn
\end{eqnarray}
Since the asymptotic parameters are real numbers, the range of momentum to mass ratio $P/M$ is restricted to be
\begin{eqnarray} \label{eq:P:range}
\frac{|a-1|}{2} \leq \left| \frac{P}{M}\right| \leq \sqrt{\frac{a^2-a+1}{3}} .
\end{eqnarray}
Note that the difference between the upper bound and the lower bound of the inequality~(\ref{eq:P:range}) is positive definite except $a=-1$, in which case $P=M$.

The transformation $\mu \rightarrow -\mu$ with $q\rightarrow -q$ is equivalent to $P\rightarrow -P$.
This parameter transformation is equivalent to the usual symmetry transformation $z\rightarrow -z$.
In appendix A, we show that the parameter $q$ is an observer dependent quantity, which varies with the motion of an asymptotic observer. For an appropriate observer, we may fix the value $q$ to zero with $a=1$. As a result, $K\cos\mu$ should be a non-negative quantity for $M$ to be non-negative number in Eq.~(\ref{relation}).
This condition and $K\geq 0$ restrict the range of the parameter $\mu$ to
\begin{eqnarray} \label{eq:theta:range}
-\frac{\pi}{2} \leq \mu \leq \frac{\pi}{2} .
\end{eqnarray}%
The $\mu \rightarrow \pi/2$ limit should be taken carefully, otherwise the mass, tension, and momentum flow would be ill defined.

\section{Metric from the point of view of comoving observer}
In this section, we analyze the geometry of the new stationary solution with metric~(\ref{Sol3}). We study the Kretschmann invariant, the coordinates (boost) transformations, and then, shows that the parameter $q$ can be fixed to zero without loss of generality.

Note that the coordinates~(\ref{Sol3}) is ill-defined at $\rho=K$ surface. Therefore, we should inspect the singular property of the surface by observing the Kretschmann invariant and the geodesic motions.
The Kretschmann invariant of the metric~(\ref{Sol3}) is given by
\begin{eqnarray}\label{eq:Kretschmann}
&&R_{\mu\nu\rho\sigma}R^{\mu\nu\rho\sigma} =\frac{64 K^6 \left(\frac{\rho+K}{\rho-K}\right)^{-\frac{8
}{\sqrt{3}\cos \mu}} \rho ^6}{3
   (K-\rho )^8 (K+\rho )^8} \left[9
    \left(1+\frac{\rho ^4}{K^4}\right)\right.\\
 &&\quad-\left.\frac{4\sqrt{3}(5-4\tan^2\mu)}{\cos\mu}\frac{\rho}{K}\left(
    1+\frac{\rho^2}{K^2}\right)+4\left(13-4\tan^2\mu-8\tan^4\mu
        \right)\frac{\rho^2}{K^2}\right]. \nn
\end{eqnarray}
At $\cos\mu=2/3$, the square-bracket can be factorized to be $-9(\rho/K-1)^2(\rho/K+1)^2$. The Kretschmann invariant vanishes at $\rho=K$ in this case.
For $\cos\mu >1/\sqrt{3}$ and $\cos \mu\neq 2/3$, the surface $\rho=K$ is a curvature singularity since the Kretschmann invariant diverges there.
On the other hand, for $\cos\mu\leq 1/\sqrt{3}$ the divergence of the Kretschmann invariant at $\rho=K$ surface disappears because the exponent of $(\rho-K)$ becomes a non-negative number.
Therefore, for $\cos\mu\leq 1/\sqrt{3}$ or $\cos \mu= 2/3$, the curvature singularity at $\rho=K$ disappears.

Note also that the curvature square is independent of $q$, which suggests the possibility $q$ being a gauge artifact.
To check this, we consider the boost transform along the $z-$coordinates in appendix~\ref{app:gauge} and show that $q$ can be nullified by the motion of an asymptotic observer.

The stationary metric~(\ref{Sol3}) with $q=0$ is not static for $\mu\neq 0$.
This implies that the momentum flow $P$ is not driven by the boost
transform.
Instead, it is related to the parameter $\mu$.
Note also that setting $\mu \rightarrow -\mu$ is
equivalent to the coordinate change $z\rightarrow -z$ accompanied by
$q\rightarrow -q$.
For $q=0$, therefore, the sign of $\tan\mu$ determines the direction
of momentum flow $P$.
We may set $q=0$ and fix the tension to mass ratio $a$ to one. In this sense, the boost transformation is related to the mass to tension ratio rather than the momentum along the extra-dimension, which is a unique feature of the present class II compared to the class I.

For simplicity, later in this paper, we set $q=0$ without loss of generality.
The metric with nonzero $q$ can be obtained by boosting the $q=0$ solution.
In addition, from now on, we consider only the case $\mu >0$ since the metric with negative $\mu$ can be obtained by flipping the $z$-coordinate.
The metric with $q=0$ now becomes
\begin{eqnarray} \label{eq:met:osc}
ds^2&=& D^{-\frac{2}{\sqrt{3}\cos\mu}}
    \left[\cos2\Upsilon(-dt^2+dz^2)
    +2\sin2\Upsilon\, dt dz\right]+G(\rho)(d\rho^2+d\Omega^2_{(2)})\\
    &=&D^{-\frac{2}{\sqrt{3}\cos\mu}}
    \left[-\left(\cos\Upsilon \,dt-\sin\Upsilon \,dz \right)^2
    +\left(\sin\Upsilon \,dt+\cos\Upsilon \,dz\right)^2 \right]
    +G(\rho)(d\rho^2+d\Omega^2_{(2)})
    \nn,
\end{eqnarray}
where the argument $\Upsilon$ of the cosine and sine functions stands for
\begin{eqnarray} \label{upsilon}
\Upsilon(\rho)=\tan\mu\log D(\rho),
\end{eqnarray}
which monotonically increases from zero to infinity as $\rho$ decreases from infinity to $K$.
The relation~(\ref{upsilon}) between $\rho$ and $\Upsilon$ can be inverted for nonzero $\mu$:
\begin{eqnarray} \label{Upsilon}
\rho &=& \coth \left(\frac{\Upsilon}{2\tan\mu}\right)\, K.
\end{eqnarray}%

The timelike 1-form field $\omega^0 = D^{-\frac{1}{\sqrt{3}\cos\mu}}(\cos\Upsilon \,dt-\sin\Upsilon \,dz)$ and the spacelike 1-form field $ \omega^4 = D^{-\frac{1}{\sqrt{3}\cos\mu}}(\sin\Upsilon \,dt+\cos\Upsilon \,dz)$ lives in a Minkowski-like metric. The 1-forms rotate as $\Upsilon$ increases with respect to the coordinates 1-forms $(dt,dz)$.

\begin{figure}
  \includegraphics[width=.6\linewidth]{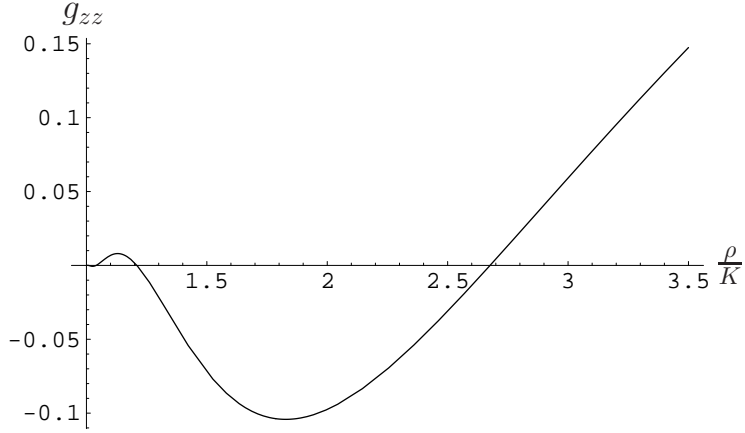}\\
  \caption{$g_{zz}=-g_{tt}$ as a function of $\rho$. In this figure, we use $\tan\mu=1$ and $K=1$. Indefinitely many roots of $g_{tt}=0$ are accumulated near $\rho=K$.}\label{gtt}
\end{figure}
The metric component $g_{tt}=-g_{zz}$ changes sign in surfaces given by
\begin{eqnarray} \label{rho:n}
 \rho_n&=&\coth(\alpha_n)\, K
    ,\quad n=0,1,2,\cdots; \quad
 \alpha_{n} =\frac{\pi}{4|\tan\mu|}
    \left(n+\frac12\right).
\end{eqnarray}
The sign of $g_{tt}$ is negative definite only in
the region,
\begin{eqnarray} \label{rho:range-}
\mathcal{R}_-= \{ {\cal E}|
 \rho >\rho_0, ~~\mbox{or }
    \rho_{2n} < \rho < \rho_{2n-1},\quad n=1,2,\cdots \},
\end{eqnarray}
and is positive definite in the region
\begin{eqnarray} \label{rho:range+}
\mathcal{R}_+ =\{{\cal E}|
    \rho_{2n+1} < \rho < \rho_{2n},\quad n=0,1,2,\cdots \}\,,
\end{eqnarray}
where ${\cal E}\equiv(t,\rho,\theta,\phi,z)$ denotes an event in this spacetime.
Therefore, at an event $\mathcal{E}\in \mathcal{R}_+$, the coordinate $t$ becomes spacelike and $z$ becomes timelike.
The asymptotic region with $\rho\geq \rho_0$ is in ${\cal R_-}$.
The repeated sign change of $g_{tt}$ is also supported by the rotational property of the 1-form basis $(\omega^0,\omega^4)$.
Note that indefinitely many $\rho_n$ are accumulated around $\rho=K$. This makes the analysis of the properties of the $\rho=K$ surface be not easy at the present form of the metric. This defect will be resolved in the next section.

The component $|g_{tz}|$ does not vanish at the surface $\rho=\rho_n$ where $g_{tt}=0=g_{zz}$.
We investigate the properties of the surface by series expanding the metric around $\rho=\rho_n$ and analyzing the radial null geodesics around there.
Ignoring the angular coordinates, the metric around $\rho=\rho_n$ takes
the form:
\begin{eqnarray} \label{met:n}
ds^2 \simeq 2(-)^n D^{-\frac{2}{\sqrt{3}\cos\mu}}(\rho_n)
dt dz+ G(\rho_n)d\rho^2 .
\end{eqnarray}
For even(odd) $n$, the direction with increasing $t+z~(t-z)$ becomes a timelike direction. The metric governing the motion along this timelike direction can be obtained by replacing $(-)^ndt dz\rightarrow -d\bar t^2 + d\bar z^2$.
After the rescalings $\sqrt{2}D^{-\frac{1}{\sqrt{3}\cos\mu}}(\rho_n)\bar t \rightarrow \bar t$ and $\sqrt{G(\rho_n)}\rho \rightarrow \rho$, we get a Minkowski-like metric in the space of $(\bar t,\rho)$.
Therefore, the geodesic on this two dimensional space $(\bar t,\rho)$ is a straight line which is free to move along the radial direction.
The only surface where this transformation is ill defined is the surface $\rho=K$ since the coordinates transformation can be singular.
Therefore, the surface $\rho=\rho_n$ has no physical significance.

The unique feature of the present metric, which says that the boost symmetry cannot be used to remove the momentum flow $P$, can be understood as follows:
In an event ${\cal E} \in {\cal R}_+$, the roles of mass and momentum are swaped. The mass $M$ and the momentum flow $P$, which is related to the translational invariance of $t$ and $z$, exchange their roles since the Killing coordinates $t$ and $z$ exchange their role as a time and a space.
The boost along the $z$-direction may change the position of the surface $\rho=\rho_n$ but can not alter the fact that there exist such ``role changing surface". In this sense, as far as we consider the solution class II, the boost can not make the momentum vanish.

\section{Causal structure}
In this section, we investigate the causal structure of the stationary metric~(\ref{eq:met:osc}) by studying spatial distance to $\rho=K$ surface and radial null geodesic motions.
Since $g_{tt}$ indefinitely oscillates as $\rho\rightarrow K$, the metric itself~(\ref{eq:met:osc}) is inconvenient to analyze its causal structure.
So, we introduce a series of coordinates transformations which changes the metric into a better form.
Then, we check the null geodesic motion of the metric to understand the causal structure.
As a result, we show that the metric describes a wormhole, a black string, and a naked singularity for parameter ranges $\cos\mu\leq 1/\sqrt{3},~1/\sqrt{3}<\cos\mu\leq \sqrt{3}/2,$ and $\cos \mu>\sqrt{3}/2$, respectively.

\subsection{spatial distance to $\rho=K$}
It is interesting to obtain the spatial distance from $\rho=K+\epsilon$
to $\rho (> K+\epsilon)$ for constant $t$, $z$, and angular coordinates, where we assume $\epsilon \ll K$ being a small number.
The spatial distance is
\begin{eqnarray} \label{distance}
l &=& \int^\rho_{K+\epsilon} d\rho \sqrt{G(\rho)} \nn \\
 &\simeq& \left\{\begin{tabular}{ll}
 $\dsp
    \frac{K\,2^{\frac{2}{\sqrt{3}\cos\mu}}}{
        \frac{1}{\sqrt{3}\cos\mu} - 1}
 \left[\left(\frac{\epsilon}{K}\right)^{2(1-\frac{1}{\sqrt{3}\cos\mu})}
    -\left(\frac{\rho}{K}-1\right)^{2(1-\frac{1}{\sqrt{3}\cos\mu})}
    \right] $, &
    $\dsp \cos\mu \neq \frac{1}{\sqrt{3}}$, \vspace{1mm}\\
 $\dsp 8\log\frac{\rho-K}{\epsilon}-4\log\frac{\rho}{K}+
    \frac{\rho}{K}+\frac{K}{\rho}-2 $\,, &
    $ \dsp \cos\mu = \frac1{\sqrt{3}}$\,. \\
\end{tabular}\right.
\end{eqnarray}
As $\epsilon \rightarrow 0$, the distance $l$ diverges for $\cos\mu
\leq 1/\sqrt{3}$ and takes a nonvanishing finite value for $\cos \mu > 1/\sqrt{3}$.
The behavior of the spatial distance for $\cos \mu \leq 1/\sqrt{3}$ shows an interesting difference from that for the usual black hole whose event horizon is in a finite distance from outside observers.
This difference is important to understand the nature of the $\rho=K$ surface.

\subsection{Geodesics motions around $\rho=K$}
Since the coordinate $(t,z)$ is inappropriate to analyze the geometry near $\rho=K$, we introduce a coordinates transformation, which makes the causal structure of the metric becomes apparent.
The coordinates rotation of $(t,z)$,
\begin{eqnarray} \label{tz:coor}
\left(\begin{tabular}{c} $t'$\\ $z'$\\ \end{tabular}\right)=
\left(\begin{tabular}{cc} $\dsp \cos\Upsilon ~$&
    $\dsp -\sin\Upsilon$
    \vspace{.1cm} \\
    $\dsp \sin\Upsilon~$& $\dsp \cos\Upsilon$\\
\end{tabular}\right)~
    \left(\begin{tabular}{c} $t$\\ $z$\\ \end{tabular}\right),
\end{eqnarray}
leads the metric~(\ref{eq:met:osc}) into the form
\begin{eqnarray} \label{met:newcoo}
ds^2&=& D^{-\frac{2}{\sqrt{3}\cos\mu}}
    \left(-d{t'}^2+d{z'}^2 - 2d\Upsilon\, d(z't')
    \right)+G(\rho)d\Omega^2_{(2)} \\
 &+& D^{-\frac{2}{\sqrt{3}\cos\mu}} \left({t'}^2-
 {z'}^2+ 2B^2(\Upsilon) \right)\,d\Upsilon^2  .
    \nn
\end{eqnarray}
The function $B(\Upsilon(\rho))$, which governs the near horizon behavior of the geodesics motion, is
\begin{eqnarray} \label{eq:B}
B(\Upsilon(\rho))&\equiv& \frac{D^{\frac{1}{\sqrt{3}\cos\mu}}(\rho)
    \sqrt{G(\rho)}}{\sqrt{2}\left|\frac{d\Upsilon(\rho)}{d\rho}\right|} \\
&=& \frac{\sqrt{2}K e^{\frac{\sqrt{3}\Upsilon}{\sin\mu}}}
    {\tan\mu\sinh^2\frac{\Upsilon}{\tan\mu}}
    = \frac{\sqrt{2}K}{4\tan\mu}\frac{\rho^2}{K^2}
    \left(1+\frac{K}{\rho}\right)^{2+\frac{\sqrt{3}}{\cos\mu}}
    \left(1-\frac{K}{\rho}\right)^{2-\frac{\sqrt{3}}{\cos\mu}}. \nn
\end{eqnarray}
For $\cos\mu <\sqrt{3}/2$ the function $B$ have a nonzero minimum at
$\rho=\rho_c$,
\begin{eqnarray} \label{eq:rhoc}
B(\Upsilon(\rho_c)) &=& \frac{K}{\sqrt{2}\sin2\mu}
    \left(\sqrt{3}+2\cos\mu\right)^{1+\frac{\sqrt{3}}{2\cos\mu}}
    \left(\sqrt{3}-2\cos\mu\right)^{1-\frac{\sqrt{3}}{2\cos\mu}}
\nn\, ,\\
\rho_c &=& \left(\frac{\sqrt{3}}{2\cos\mu}+\sqrt{\frac{3}{4\cos^2\mu}-1}
    \right)K .
\end{eqnarray}
For $\cos\mu >\sqrt{3}/2$, $B(\Upsilon(\rho))$ monotonically decreases to zero as $\Upsilon \rightarrow \infty~(\rho\rightarrow K)$.
For $\cos\mu=\sqrt{3}/2$, $B(\Upsilon(\rho))$ monotonically decreases to a finite value $4\sqrt{6}K$ as $(\rho\rightarrow K)$.
This behavior of $B$, in fact, governs the geodesic motion at $\rho \sim K$.
\begin{figure}
  \includegraphics[width=.6\linewidth]{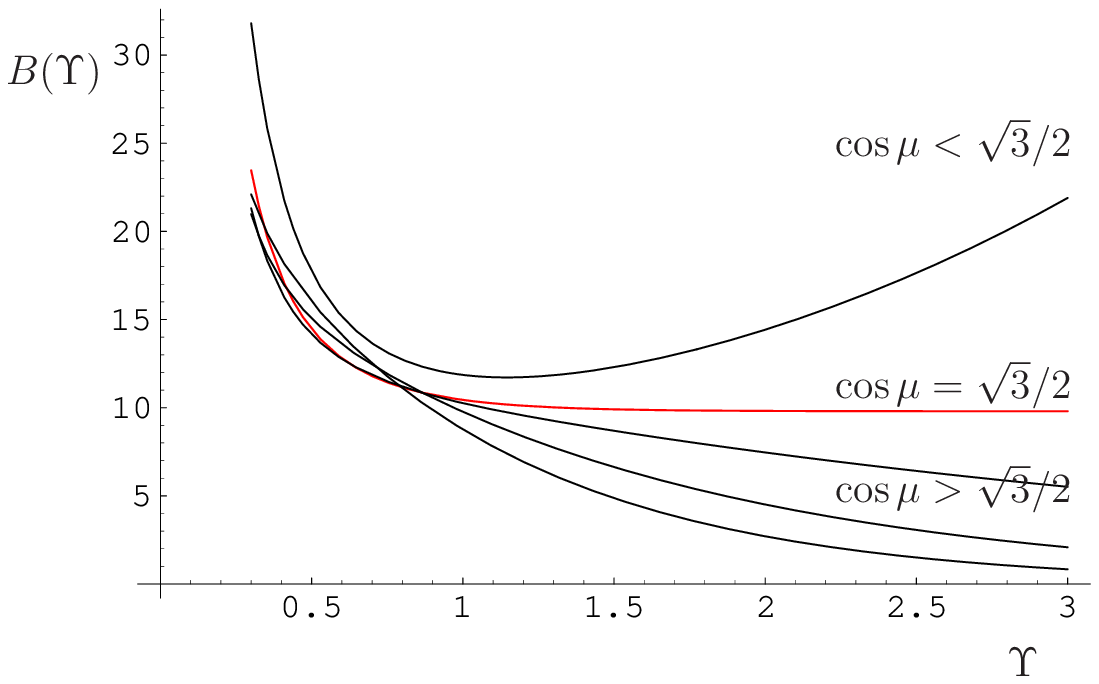}\\
  \caption{{\it color online}, $B(\Upsilon)$.\\
  As $\rho\rightarrow K$, the behavior of $B(\Upsilon)$ is divided into three classes. It diverges ($\cos\mu < \sqrt{3}/2$), goes to zero ($\cos\mu>\sqrt{3}/2$), or takes a finite value ($\cos\mu=\sqrt{3}/2$). On the other hand, the limiting behavior as $\rho \rightarrow \infty$ are the same.
  }\label{Bfn}
\end{figure}

The metric~(\ref{met:newcoo}) is dependent on $t'$ and $z'$ explicitly.
Therefore, the coordinates rotation~(\ref{tz:coor}) hides the beautiful symmetry of the translational invariances along the fifth coordinate and time.
Notably, the lines $t'=\pm z'$ ($\rho,\theta,\phi$ being fixed) describe lightlike geodesics.
If one view only the $\Upsilon=$ constant surface, the spacetime is just described by the space $R^2\times S^2$ where $R^2$ has the Minkowski signature.

In addition, the spacelike vector $\left(\frac{\partial}{\partial \rho}
\right)^{a}_{(tz)}$, where the index $(tz)$ implies the vector is defined in the unprimed coordinates, is related to the vectors in the primed coordinates by
\begin{eqnarray} \label{eq:vectorrel}
 \left(\frac{\partial}{\partial \rho}
    \right)^{a}_{(tz)}= \left(\frac{\partial}{\partial \rho}
    \right)^{a}_{(t'z')}+\frac{d\Upsilon}{d\rho}\left(t'
        \frac{\partial}{\partial z'}-z'\frac{\partial}{\partial t' }
        \right)^a .
\end{eqnarray}
Even though the vector $\left(\frac{\partial}{\partial \rho}
\right)^{a}_{(tz)}$ is spacelike always, its primed coordinates form $\left(\frac{\partial}{\partial \rho}\right)^{a}_{(t'z')}$ can be timelike for certain region of spacetime satisfying ${z'}^2>{t'}^2+2B^2$.

\begin{figure}
  \includegraphics[width=.4\linewidth]{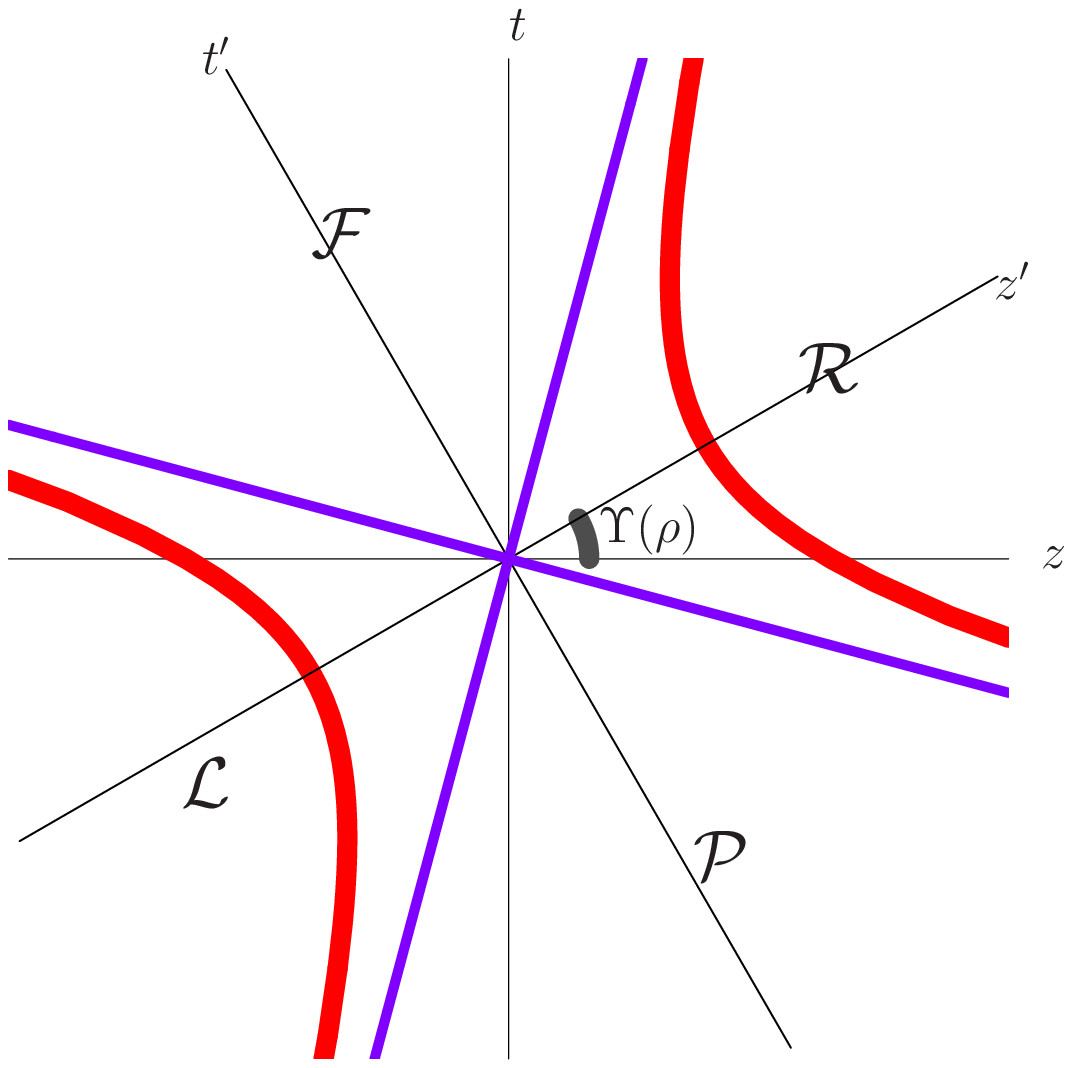}\\
  \caption{color online. Relation of the two coordinates $(t,z)$ and $(t',z')$. \\
  The thick curve denotes the ${z'}^2={t'}^2+2 B^2(\rho)$ surface.
  The thick orthogonal lines denote lightcones bifurcating from the origin. For a given radial coordinate $\rho$, the axes $(t',z')$ rotates with the angle $\Upsilon$ with respect to $(t,z)$. The lightcone is designed with respect to this primed coordinates. We are mainly interested in the regions inside the red hyperbola containing the origin $(t',z')=(0,0)$ in the present section. }\label{fig:coo:prime}
\end{figure}

The metric component $g_{\rho\rho}'$ is positive for $z'^2< t'^2 +2 B^2(\rho)$  and is negative for $z'^2> t'^2 +2 B^2(\rho)$, where the surface ${z'}^2= {t'}^2+2B^2$ is denoted by a red thick hyperbola in Fig.~\ref{fig:coo:prime}.
In the region of spacetime satisfying $z'^2> t'^2 +2 B^2(\rho)$
the vector $\left(\frac{\partial}{\partial \rho}\right)^\mu$ becomes time-like and there is a frame dragging along $z'$-coordinate due to the presence of $dz'd \Upsilon$ term.
Therefore, we should include the dynamics along the $z'$-coordinate to analyze the geometry there and the analysis of the geodesic equation with this metric is not easy. In this sense, this metric~(\ref{met:newcoo}) is not appropriate to understand the full structure of the spacetime since the motion along $z'$-coordinates are mixed up with that of along $\Upsilon$.
For the region $z'^2< t'^2 +2 B^2(\rho)$, both of the vectors $\left(\frac{\partial}{\partial z'}\right)^\mu$ and $\left(\frac{\partial}{\partial \rho}\right)^\mu$ are spacelike. As $\rho \rightarrow K$, the entire spacetime satisfies this inequality $z'^2< t'^2 +2 B^2(\rho)$ for $\cos\mu <\sqrt{3}/2$ since $B^2(K)$ diverges.

In the present section, we discuss the geometry only in the region satisfying $t'\geq |z'| $ (region ${\cal F}$ in Fig.~\ref{fig:coo:prime}), in which the geodesic motion is transparent. We present the analysis for other regions at appendix B.
To visualize the near horizon geometry in ${\cal F}$, we perform coordinates transform,
\begin{eqnarray} \label{eq:prime:tauy}
y= t'z', \quad \quad \tau= ({t'}^2+{z'}^2)/2\, ,
\end{eqnarray}
and get the transformed metric
\begin{eqnarray} \label{eq:met:F}
ds^2= D^{-\frac{2}{\sqrt{3}\cos\mu}}\left[
    \frac{-d \tau^2+(d y- 2\sqrt{\tau^2-y^2}\, d\Upsilon)^2}{2\sqrt{\tau^2-y^2}}
    +2 B^2(\Upsilon)d \Upsilon^2\right]+G(\rho) d\Omega^2_{(2)} ,
\end{eqnarray}
where ${t'}^2-{z'}^2 =2\sqrt{\tau^2-y^2}$. Note that this metric is applicable only on the one quarter of the spacetime satisfying $t'> |z'|$, which boundary will be given by $\tau=|y|$.
The eventual coordinates transformation from $(t,z,\Upsilon)$ to $(\tau,y,\Upsilon)$, using complex parametrization, $z+it = r e^{i\alpha}$, is given by
\begin{eqnarray} \label{coo:F}
\tau &=& \frac{r^2}{2} , \quad y =\frac{r^2}{2}\sin 2(\alpha-\Upsilon)\,,
\end{eqnarray}
where the region $\mathcal{F}$ is mapped to
\begin{eqnarray} \label{RF}
\mathcal{F} &=& \{{\cal E}=(t,\rho,\theta,\phi,z)| \Upsilon(\rho)+\frac{\pi}{4}<\alpha< \Upsilon(\rho)
   + \frac{3\pi}{4}\} \,. \nn
\end{eqnarray}
In the asymptotic region, $\Upsilon =0$. Therefore, the region ${\cal F}$ corresponds to the region $t\geq |z|$ asymptotically.
The region rotates as $\rho$ decreases keeping the timelikeness with respect to the origin $(t,z)=(0,0)$.
For the new metric~(\ref{eq:met:F}), the coordinate $\tau$ is a time and the coordinate $y$ is a space.

Let us study the near horizon geometry by investigating the geodesic motion of the metric~(\ref{eq:met:F}).
To study the geodesic motion in the radial $\rho$ coordinate, we need the formula:
\begin{eqnarray} \label{eq:d rho:d t}
\dot \rho = -\frac{K }{2\tan\mu}\frac{\rho^2}{K^2}\left(1-\frac{K^2}{\rho^2}\right)
    \dot\Upsilon \,,
\end{eqnarray}
where $\dot \rho$ implies derivative of $\rho$ with respect to $\tau$.
From Eq.~(\ref{eq:met:F}), we obtain the  radial null geodesics equation,
\begin{eqnarray} \label{eq:gedesic}
(\dot y-2\sqrt{\tau^2-y^2} \, \dot \Upsilon )^2+
    4\sqrt{\tau^2-y^2} B^2(\Upsilon) \dot \Upsilon^2 =1.
\end{eqnarray}
The geodesic equation~(\ref{eq:gedesic}) restricts the range of $\dot \Upsilon$ to
\begin{eqnarray} \label{range:2}
-\frac{1}{2(\tau^2-y^2)^{1/4} B} \leq &\dot \Upsilon  &\leq
    \frac{1}{2(\tau^2-y^2)^{1/4} B}\,,
\end{eqnarray}
where the equalities hold when $\dot y= 2\sqrt{\tau^2-y^2} \dot \Upsilon$.
The limiting behavior of the function $\lim_{\Upsilon \rightarrow \infty} B(\Upsilon)$ is characterized by the sign of $\cos\mu-\sqrt{3}/2$.
For $\cos\mu>\sqrt{3}/2$, $B$ vanishes and the maximum value of $\dot \Upsilon$ in the limit diverges.
For $\cos\mu<\sqrt{3}/2$, $B$ diverges and the velocity $\dot \Upsilon$ in the limit vanishes.

This limiting behavior of the velocity, in fact, determines the nature of the surface $\rho=K$.
To show this, we rewrite Eq.~(\ref{range:2}) in terms of the velocity $\dot\rho$ and then integrate the equation to get the escaping time from the $\rho=K$ surface.
Using Eq.~(\ref{eq:d rho:d t}), the velocity $\dot \rho$ is restricted to
\begin{eqnarray}
-\frac{\left(1+\frac{K}{\rho}\right)^{-1-\frac{\sqrt{3}}{
    \cos\mu}}
    \left(1-\frac{K}{\rho}\right)^{-1
    +\frac{\sqrt{3}}{\cos\mu}}}{\sqrt{2} (\tau^2-y^2)^{1/4}}\leq
    &\dot \rho & \leq\frac{\left(1+\frac{K}{\rho}\right)^{-1-\frac{\sqrt{3}}{
    \cos\mu}}\left(1-\frac{K}{\rho}\right)^{-1+\frac{\sqrt{3}}
    {\cos\mu}}}{\sqrt{2} (\tau^2-y^2)^{1/4}} ,\nn
\end{eqnarray}
The maximal velocity of radial null geodesics around $\rho\sim K$ is determined by  $(\sqrt{3}/\cos\mu -1)$, the exponent of $(1-K/\rho)$.
Since this exponent is positive definite, the allowed velocity, $\dot \rho$, is restricted to zero at $\rho=K$, which is a typical feature of horizon.

We next integrate out the geodesic equation to calculate how much time does it take a light to escape from $\rho=K+\epsilon$ to $\rho$.
For radially outgoing geodesics, the maximal radial velocity is achieved for $\dot y = 2\sqrt{\tau^2-y^2} \dot \Upsilon$.
In this case, $\dot \rho \propto \tau^{-1/2} (\rho-K)^{\frac{\sqrt{3}}{\cos\mu} -1}$.
Integrating the equation, we get the escaping time
\begin{eqnarray} \label{Delta:t1}
\Delta \tau \propto (\rho -K)^{4-\frac{2\sqrt{3}}{\cos\mu}} ~\mbox{ for
    }\cos\mu \neq \frac{\sqrt{3}}{2},
\end{eqnarray}
and for $\cos\mu = \frac{\sqrt{3}}{2}$, we have
\begin{eqnarray} \label{Delta:t2}
\Delta \tau \propto \log (\epsilon).
\end{eqnarray}
The escaping time $\Delta \tau$ diverges for $\cos\mu \leq \sqrt{3}/2$,  therefore no information can get out of the surface $\rho =K$ with $\tau> |y|$.
Similarly, for ${\cal P}$ in Fig.~\ref{fig:coo:prime}, no radial null geodesics can get out of the $\rho=K$ surface if $\cos\mu \leq \sqrt{3}/2$.
On the other hand, the surface $\rho=K$ is naked singular if $\cos\mu >\sqrt{3}/2$ since the Kretschmann invariant diverges there (See Eq.~(\ref{eq:Kretschmann})) and the time getting out of the surface is finite (\ref{Delta:t1}).

The causal structure for the region $\mathcal{R}$ and $\mathcal{L}$ are studied in the appendix~\ref{sec:appB}. The radial null geodesics in the region $|z'|^2< {t'}^2+2B^2$ at $\rho=K$ cannot escape the surface.
For $\mu > \pi/6$, the value $B(K)$ diverges and therefore, all events at the surface $\rho=K$ are satisfied with the inequality $|z'|^2< {t'}^2+2B^2$. Therefore, all geodesics starting from the surface cannot escape the surface in a finite time.

The unit one-form denoting the advances of time is
$\omega^{t}= D^{-\frac{1}{\sqrt{3}\cos\mu}}( dt'+z' d\Upsilon) $ as one sees in Eq.~(\ref{met:newcoo}).
For region with $|z'|> \sqrt{{t'}^2+2B^2}$, the vector $\left(\frac{\partial}{\partial \rho}\right)^{a}_{(t'z')}$ becomes timelike and the situation becomes nontrivial.
Especially, the surface $\rho=K$ with $z'<-\sqrt{{t'}^2+2B^2(K)}$ becomes a naked singularity.
For $\cos\mu=\sqrt{3}/2$, $B(K)$ takes a finite value.
Therefore, the singularity at $\rho=K$ will be seen by the outside observer through the trajectory in the region $z'<-\sqrt{{t'}^2+2B^2}$. However, the information on the surface $\rho=K$ with $t'>z'$ still cannot escape.
In this sense, the solution with $\cos \mu=\sqrt{3}/2$ is a mixture of naked singularity and black string.

Summarizing the results in this section, the surface $\rho=K$ forms an event horizon for the parameter range $1/\sqrt{3} < \cos \mu < \sqrt{3}/2$.
For $\cos \mu=\sqrt{3}/2$, the surface $\rho=K$ with $t'>z'$ ($z'<-\sqrt{{t'}^2+2B^2}$) forms an event horizon (a naked singularity). Therefore, this solution is a kinds of interwinding solution of naked singularity and black string.
If one compactify the fifth coordinate so that it contains only the regions satisfying $|z'| \leq B$, the solution becomes black string.
For $\cos \mu>\sqrt{3}/2$, the surface $\rho=K$ is a naked singularity.
For $\cos \mu \leq 1/\sqrt{3}$, the surface $\rho=K$ becomes a spatial infinity. As we will show in the next section, the region $\rho\sim K$ becomes another asymptotic region of space connected with the asymptotic region $\rho\rightarrow \infty$ through a wormhole throat.
In this sense, the solution with $\cos \mu \leq 1/\sqrt{3}$ describes a wormhole geometry. Solution of similar wormhole geometry was found Cl\'{e}ment~\cite{clement}, which in fact corresponds to the generalization of class III.

\section{Geometric properties of the solution}
In this section, we study the geometrical properties of the solution such as the volume of $S^2$ sphere for a given $\rho$, $t$ and $z$, and the proper length corresponding to the fifth coordinate distance.
We concentrates on qualitative behaviors of the quantities and we use the coordinates system~(\ref{eq:met:osc}) in which the Killing symmetries of $z$ and $t$ are evident except for the calculation of the distance along the fifth coordinate.

\subsection{Area of $S^2$ sphere}

The behavior of the area for the $S^{2}$ sphere in $z=$ constant spacelike surface is
\begin{eqnarray} \label{A}
A_{2}(\rho) &=& 4\pi(\sqrt{G}\rho )^{2}\\
   &=&4\pi
   \rho^{2}\left(1-\frac{K}{\rho}\right)^{2-\frac{4}{
        \sqrt{3}\cos\mu}} \times
    \left(1+\frac{K}{\rho}\right)^{2+\frac{4}{\sqrt{3}\cos\mu} }. \nn
\end{eqnarray}
Note that $2-4/(\sqrt{3}\cos\mu)$, the exponent of $(1-K/\rho)$, is always negative definite. As $\rho$ approaches to $K$ from infinity, the area decreases to its minimum value at
\begin{eqnarray} \label{eq:rhom}
\rho=\rho_m(\mu) =\left(\frac{2}{\sqrt{3}\cos\mu}+
    \sqrt{\left(\frac{2}{\sqrt{3}\cos\mu}\right)^2-1}\right)K \geq
    \sqrt{3} K,
\end{eqnarray}
then bounces up to infinity instead of decreasing down to zero monotonically. $\rho_m(\mu)$ is an increasing function of $\mu (\geq 0)$ and diverges as $\mu \rightarrow \pi/2$.
At $\rho=K$, the surface area $A_{2}(K)$ diverges for all value of
$\mu$.

Considering the two facts that the infinite distance to $\rho=K$ surface (\ref{distance}), and the infinite area of the $S^2$ sphere, we may identify that there is infinite large volume of space near $\rho=K$ surface for $\cos\mu \leq 1/\sqrt{3}$. Since the curvature square and curvature tensors vanishes for $\cos \mu <1/\sqrt{3}$ the spacetime becomes an another asymptotically flat region connected with the region $\rho \rightarrow \infty$ through the wormhole throat at $\rho=\rho_m$.
Therefore, the geometry around $\rho=\rho_m$, becomes a kind of wormhole throat for $\cos \mu \leq 1/\sqrt{3}$.

For the case of $\cos\mu > 1/\sqrt{3}$, even though the $S^2$ area of $\rho=K$ surface is divergent, the distance to the surface is finite.
In addition the curvature tensor does not vanish.
Thus, we can not interpret the $\rho=K$ surface as an asymptotic region.

\subsection{Proper length for unit fifth coordinate distance}

The spatial geometry along the fifth coordinate may be understood by considering the proper length along the fifth coordinates with $\rho=$ constant and the angular coordinates being fixed. Because of the high oscillatory nature of the metric component $g_{zz}$ in Eq.~(\ref{eq:met:osc}) around the horizon, we cannot use it to measure the proper distance along the direction. Instead, we measure the proper distance corresponding to unit $z'$ coordinates from Eq.~(\ref{met:newcoo}). Since the coordinates $z$ and $z'$ are the same asymptotically and the order of $g_{zz}$ is the same as that of $g_{z'z'}$ up to oscillatory nature, we use the second quantity to measure the proper distance:
\begin{eqnarray} \label{length}
&&L(\rho) \propto \int_{z'}^{z'+1}\sqrt{g_{z'z'}} dz' =
    D^{-\frac{1}{\sqrt{3}\cos\mu}} .
\end{eqnarray}
The distance monotonically increases from zero to unity as $\rho$ increases from $K$ to infinity.
Therefore, any non-zero coordinate distance along the fifth coordinate takes zero proper distance at the horizon.

\subsection{Horizon area}

The physical spacelike area of the $\rho=$ constant surface per unit length of $z'$ are given by the product of the $S^2$ area and the proper length along the fifth coordinate:
\begin{eqnarray} \label{eq:S}
A(\rho) = A_{2}(\rho)\times L(\rho)
    & \propto &4\pi
   \rho^{2}\left(1-\frac{K}{\rho}\right)^{2-\frac{\sqrt{3}}{\cos\mu}} \times
    \left(1+\frac{K}{\rho}\right)^{2+\frac{\sqrt{3}}{\cos\mu}}
    \\
 &\simeq&
     \left(1-\frac{K}{\rho}\right)^{2-\frac{\sqrt{3}}{\cos\mu}} , \nn
\end{eqnarray}
where the second line denotes its behavior around $\rho\simeq K$.
At $\rho=\infty$, this area diverges because of the $\rho^2$ term.
For $\cos\mu <\sqrt{3}/2$, as $\rho$ approaches to $K$ from infinity, the area decreases to its minimum value at
\begin{eqnarray} \label{eq:rhom}
\rho=\rho_{min}(\mu) =\left(\frac{2}{\sqrt{3}\cos\mu}+
    \sqrt{\left(\frac{2}{\sqrt{3}\cos\mu}\right)^2-1
    -\frac{2}{\sqrt{3}\cos\mu}}\right)K ,
\end{eqnarray}
then bounces up to infinity instead of decreasing down to zero monotonically.
$\rho_{min}(\mu)$ is an increasing function of $\mu$ with $\rho_{min}(\pi/6)=K$.

For $\cos\mu > \sqrt{3}/2$, the area $A(\rho)$ is a monotonically increasing function of $\rho\geq K$ and vanishes at $\rho=K$.
For $\cos\mu=\sqrt{3}/2$, the area $A(\rho)$ is a monotonically increasing function of $\rho\geq K$ and takes a nonzero finite value at $\rho=K$.

If we interpret this area as the entropy of the black string, the entropy of the black string diverges (vanishes) for $ \cos \mu >\sqrt{3}/2~(\cos \mu<\sqrt{3}/2)$. A finite entropy of the black string is given for $\cos \mu=\sqrt{3}/2$.

\section{Summary and discussion}
By using the analytic continuation procedure of coefficients after boosting static solution, we have shown that there are three classes of stationary string-like solutions of the Einstein equation in (4+1) dimensions.
The class I is what we usually expect to exist, the solution obtained from boosting the static one.
The class III is a kind of wormhole-like solution known in Ref.~\cite{chodos}, which we do not discuss in this paper.
We are mainly interested in the class II, which intertwining the two classes I and III because it shows continuous spectrum of naked singularity, black string, and wormhole solutions in parameter space.
The class II is characterized by three parameters $(K, \mu; q)$ related to the mass $M$, tension $\tau$, and the momentum flow $P$ along the fifth coordinate, where $q$ varies under the boost of observers. Restricting ourselves to $q=0$ case, we analyzed the causal structure of the metric to show that it describes a black string for the range of parameter $1/\sqrt{3}<\cos\mu\leq \sqrt{3}/2$ and a wormhole for $\cos \mu\leq 1/\sqrt{3}$.
The momentum to mass ratio of the solution is restricted by the value of the tension to mass ratio according to  Eq.~(\ref{eq:P:range}). Interestingly, solution with zero momentum does not exist in this class except for the case of $\cos\mu=1$.

We briefly summarize the geometric properties of the solution in table I.
\begin{table}
\begin{tabular}{|c|c|c|c|c|c|c|c|}
\hline
$\cos\mu $ & $ 1 \sim$  & $\frac{\sqrt{3}}2$
    & $\sim$ & $\frac23$ & $\sim$ &
    $\frac1{\sqrt{3}}$ &
    $\sim 0 $ \\ \hline
$\left|\frac{P}{M}\right| $ & $ 0 \sim$  & $\frac1{2\sqrt{3}}$
    & $\sim$ & $\frac{\sqrt{5}}{3\sqrt{3}}$ & $\sim$ &
    $\frac{\sqrt{2}}{3}$ &
    $\sim \infty $ \\ \hline
Distance to $\rho=K$
    &  finite & finite  & finite & finite & finite &
    $\infty$ &$\infty $ \\ \hline
Kretschman invariant
    &$\infty $ &$\infty$ & $\infty $ &$0$ &$\infty$ & finite & $ 0$
        \\    \hline
Nature of $\rho=K$ &naked singularity &  horizon & horizon & horizon &
        horizon &$i_0$ & $i_0$ \\ \hline
area of $S^2$ & $\infty $ & $\infty$ & $\infty $ & $\infty$ &
    $\infty $& $\infty$ & $\infty$ \\ \hline
Length for unit $z$ & $0 $ & $0 $ & $0 $ & $0$ & $0 $& $0$ & $0 $
    \\ \hline
area of $\rho=K$ & $0$ &finite & $\infty$ & $\infty$  & $\infty$ & $\infty$  & $\infty$ \\ \hline
spacetime nature & naked singularity & \multicolumn{4}{c|}{black string} &
     \multicolumn{2}{c|}{wormhole}  \\ \hline
\end{tabular}\\
\caption{Geometric properties of the solution}
\end{table}
For $\cos\mu\leq 1/\sqrt{3}$, the surface $\rho=K$ is located at spatial infinity and a wormhole throat is located at $\rho=\rho_m> K$ given in Eq.~(\ref{eq:rhom}).
For $\cos \mu= \sqrt{3}/2$, the horizon area of the black string is finite. However, the extension of the geometry to the behind of the horizon may be impossible since the Kretschmann invariant diverges at the horizon.
For $\cos \mu= 2/3$, the horizon is nonsingular and is in a finite distance from the outside observers. Therefore, this metric may have hidden spacetime behind the horizon, which will be obtained by using Kruskal-like extension.
For $\cos \mu=1/\sqrt{3}$, the Kretschmann invariant takes finite value at $\rho=K$ and the $\rho=K$ surface area diverges. In addition, the distance to the horizon diverges. Therefore, the spacetime is composed of two regions of infinite volume connected by a wormhole throat located at $\rho_m=(2+\sqrt{3})K$ in Eq.~(\ref{eq:rhom}).

The parameter $\cos\mu$ decreases as the conserved momentum increases. For $\cos \mu= \frac{\sqrt{3}}2~(\frac1{\sqrt{3}})$, using Eq.~(\ref{relation}), the momentum satisfies $P_{c}= \frac{1}{2\sqrt{3}}M~ (P_w=\frac{\sqrt{2}}{3}M)$. The subscript $c$ $(w)$ denotes the critical momentum value which divides the parameter space into regions with naked singularity and black string (black string and wormhole). For small of the momentum $|P| < P_c$, the solution is naked singular.
For momentum with $P_c\leq |P|< P_w$, the solution becomes a black string.
For large momentum flow $|P|\geq P_w$, the solution describes a wormhole geometry.
In conclusion, the metric describes a naked singularity, a black string, and a wormhole as the momentum along the $z$-coordinate increases.

Most of the present black string solutions have divergent horizon area. Interpreting the area as an entropy of the black string, this implies that the black string solution would be entropically stable.
This result may shed a new light on the stability problem of the black string solution.

Since any stable configuration of non-rotating, spherically symmetric, vacuum black string is not known yet, it was a big issue to know the final fate of cylindrically collapsing matters in higher dimensions.
The present solution might be a candidate of the final state since it is at least entropically stable. To ensure this it is important to understand the full stability (thermodynamic and metric-perturbational) of the metric.

One of an important question unanswered yet is how can we construct a solution with a compact extra dimension. Simple compactification of $z$-coordinate leads to the formation of closed timelike curves near the horizon because the Killing vector $\left(\frac{\partial}{\partial z}\right)^a$ becomes timelike around the horizon. Since we live in four dimensional spacetime now, it is necessary to obtain a four dimensional interpretation of the metric using an appropriate compactification method. We leave this subject as a further work.

\vspace{1cm}
\begin{appendix}
\section{Gauge transform} \label{app:gauge}
We study the stationary solution~(\ref{Sol3}), class II, seen
by a moving observer with velocity $v=\tanh \beta$ along the
$z-$direction.
The coordinates transformation is given by
\begin{eqnarray} \label{boost:asym}
\left(\begin{tabular}{c}
    $t'$\\ $z'$ \\\end{tabular}\right)
    &=&\left(\begin{tabular}{c}
    $\cosh \beta~~ \sinh\beta$\\ $\sinh\beta~~
        \cosh\beta$ \\ \end{tabular}\right)
\left(\begin{tabular}{c}
    $t$\\ $z$ \\\end{tabular}\right),
\end{eqnarray}
where we ignore the angular and the radial directions for simplicity.
From the point of view of this observer, the metric~(\ref{Sol3}) takes
the form:
\begin{eqnarray} \label{met:v}
ds^2_{\beta}&=& \left\{\left[g_{tt}- 2\tanh\beta \, g_{tz}
    +\tanh^2\beta \,g_{zz}\right]{dt'}^2
+\left[g_{zz}-2\tanh\beta\, g_{tz} +\tanh^2\beta
    g_{tt}\right]{dz'}^2\right. \nn\\
&+&\left.2\left[-\tanh\beta (g_{tt}+g_{zz}) +(1+\tanh^2\beta) g_{tz}\right]dt'dz'\right\}\cosh^2\beta .
\end{eqnarray}
The components of the metric~(\ref{met:v}) is
\begin{eqnarray} \label{gtt'}
g_{tt}' &=& -D^{-\frac{2}{\sqrt{3}\cos\mu}}\left(
    \cos 2\Upsilon+ 2q_\beta \sin 2\Upsilon
         \right) \, ,\\
g_{zz}' &=& D^{-\frac{2}{\sqrt{3}\cos\mu}}\left(
    \cos 2\Upsilon- 2q_\beta \sin2\Upsilon
         \right) \,, \nn\\
g_{tz}' &=&D^{-\frac{2}{\sqrt{3}\cos\mu}}\sin2\Upsilon
  \left[(1+4q^2)^{1/2}\cosh2\beta+
    2q\sinh2\beta \right]
         ,\nn
\end{eqnarray}
where $q_\beta =q\cosh2\beta+(1/4+q^2)^{1/2}\sinh2\beta$.
If we determine $\beta$ so that $q_\beta=0$, the square bracket in the third line of Eq.~(\ref{gtt'}) becomes one and the transformed metric
$g_{\mu\nu}'$ becomes the metric~(\ref{Sol3}) with $q=0$.
In this sense, the metric~(\ref{met:v}) is equivalent to the metric~(\ref{Sol3})
with $q=0$ up to the boost symmetry of the coordinate.

\section{Geodesic motions in region ${\cal R}$ }\label{sec:appB}

In this appendix, we study the null geodesic motion in the region ${\cal R}$ and ${\cal L}$ in Fig.~(\ref{fig:coo:prime}).
\begin{figure}
  \includegraphics[width=.4\linewidth]{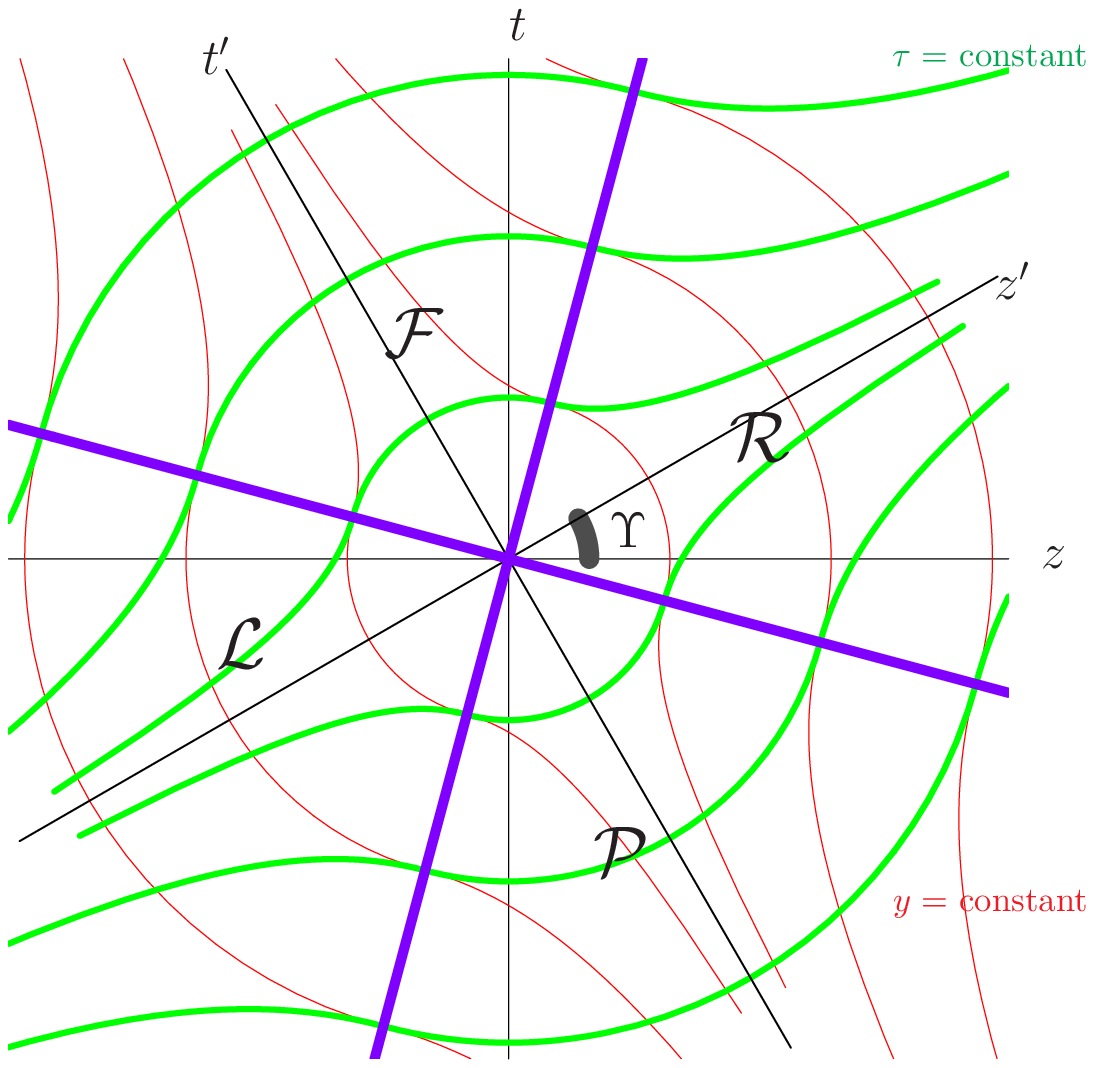}\\
  \caption{color online. Relation of three coordinates $(t,z)$, $(t',z')$, and $(\tau,y)$. \\
  The red curve is the constant $y$ surface and the green curve denotes the constant $\tau$ spacelike surface.
  The thick blue lines denote lightcones bifurcating from the origin. For a given radial coordinate $\rho$, the axes $(t',z')$ rotates by the angle $\Upsilon$ with respect to $(t,z)$. The lightcone is designed with respect to this primed coordinates. }\label{fig:coo:tauy}
\end{figure}
In the regions  ``${\cal R}$" and ``${\cal L}$", where $|z'|\geq |t'|$ (the events in region is connected with a spacelike geodesics with the origin), we use the coordinates transformation, $\tau=\pm t'z'$ and $y=\pm ({t'}^2+{z'}^2)/2$, where we use positive( negative) sign for $`` \mathcal{R}"~(`` \mathcal{L}")$ to keep the directions of timelike and spacelike coordinates and get the metric:
\begin{eqnarray} \label{eq:met:R}
ds^2_{s,\pm}= D^{-\frac{2}{\sqrt{3}\cos\mu}}
    \frac{d y^2-(d \tau \pm 2\sqrt{y^2-\tau^2}\, d\Upsilon)^2}{2\sqrt{y^2-\tau^2}}
    +G(\rho)(d\rho^2+d\Omega^2_{(2)}) ,
\end{eqnarray}
where ${z'}^2-{t'}^2 = 2\sqrt{y^2-\tau^2}$.
Coordinates singularity happens along the lightlike surface $\tau=\pm y$, however, they does not give rise to any singular behavior of the geodesics.
At an event ${\cal E}\in {\cal R}$, there happens dragging of coordinates frame along the $\rho$ coordinate.
The eventual coordinates transformation from $(t,z,\Upsilon)$ to $(\tau,y,\Upsilon)$ is, using complex parametrization $z+it = r e^{i\alpha}$, given by
\begin{eqnarray} \label{coo}
\tau &=&\pm \frac{r^2}{2}\sin 2(\alpha-\Upsilon) , \quad
 y= \pm \frac{r^2}{2} \,, \quad
    \mbox{for } \mathcal{R/L}  \,,
\end{eqnarray}
where
\begin{eqnarray} \label{RF}
\mathcal{R} &=& \{{\cal E}|~
     |\alpha-\Upsilon| \leq \frac{\pi}{4}\}  \,,\\
\mathcal{L} &=& \{{\cal E}|~
     |\alpha-\Upsilon -\pi| \leq \frac{\pi}{4} \}  \,.\nn
\end{eqnarray}
In the asymptotic region, $\Upsilon =0$. Therefore, the region $\mathcal{R}$ corresponds to the region $|t|< z$.
The region rotates counter-clockwisely as $\rho~(\Upsilon)$ decreases (increases) keeping the spacelike property of the region $\mathcal{R}$ with respect to the origin.

Since the lightcone, $t'=\pm z'$, is a coordinates singularity, we add the metric along the lightcone:
\begin{eqnarray} \label{eq:met:l}
ds_{l}^2&=& e^{-\frac{2\Upsilon}{\sqrt{3}\sin\mu}}
    \left[\mp d T
        +2 B^2
   \,d\Upsilon\right]d\Upsilon+G(\Upsilon)d\Omega^2_{(2)} ,
\end{eqnarray}
where $T= {t'}^2$.
The null geodesics satisfies
\begin{eqnarray} \label{eq:geodesic:l}
\frac{d\Upsilon}{dT} =0, ~~~~
\frac{d\Upsilon}{dT} = \pm \frac{1}{2B^2}.
\end{eqnarray}
The maximal velocity of the outgoing null geodesics to the radial direction comes from the second equation of (\ref{eq:geodesic:l}) with positive sign. This velocity vanishes for $\cos\mu \leq \frac{\sqrt{3}}{2}$ at $\rho=K$.
This is a kind of indication of horizon of the $\rho=K$ surface for $\cos\mu \leq \frac{\sqrt{3}}{2}$.

Now we study the near horizon geometry in $\mathcal{R}$ by investigating the geodesic motion of the metric~(\ref{eq:met:R}) with positive sign.
The radial null geodesic of the metric~(\ref{eq:met:R}) satisfies,
\begin{eqnarray} \label{geodesic:R}
\dot y^2= (1+2\sqrt{y^2-\tau^2} \dot \Upsilon )^2-
    4\sqrt{y^2-\tau^2}B^2 \dot \Upsilon^2\geq 0
    .
\end{eqnarray}
Solving Eq.~(\ref{geodesic:R}), the radial velocity is given by
\begin{eqnarray*} \label{dUpsilon:3}
\dot \Upsilon = \frac{1}{2(B^2-\sqrt{y^2-\tau^2})} \left[1 \pm
    \sqrt{1+ \frac{(1-\dot y^2)(B^2-\sqrt{y^2-\tau^2})}{\sqrt{y^2-\tau^2}}}
        \right].
\end{eqnarray*}
The spacetime is divided into two different regions according to the sign of $B^2-\sqrt{y^2-\tau^2}$.

Let us call the two regions by ${\cal R}_1$ and ${\cal R}_2$:
\begin{eqnarray} \label{R12}
{\cal R}_1 = \{{\cal E}\in {\cal R}| y^2< \tau^2+ B^4   \},\quad\quad
{\cal R}_2 = \{{\cal E}\in {\cal R}| y^2\geq \tau^2+ B^4   \} \,.
\end{eqnarray}
In the region ${\cal R}_1$, with $0<(y^2-\tau^2)^{1/4} < B$, the radial velocity is restricted to
\begin{eqnarray} \label{dUpsilon:2}
-\left[\frac{B}{(y^2-\tau^2)^{1/4}}+1\right]^{-1} \leq 2(y^2-\tau^2)^{1/2}\dot \Upsilon\leq \left[\frac{B}{(y^2-\tau^2)^{1/4}}-1\right]^{-1},
\end{eqnarray}
where the equality holds for $\dot y=0$.
Using Eq.~(\ref{eq:d rho:d t}), the velocity $\dot \rho$ is restricted to
\begin{eqnarray} \label{dUpsilon:2}
&&- \frac{1-K^2/\rho^2}{
   \left(1+\frac{K}{\rho}\right)^{2+\frac{\sqrt{3}}{\cos\mu}}
    \left(1-\frac{K}{\rho}\right)^{2-\frac{\sqrt{3}}{\cos\mu}}
    -\frac{2\sqrt{2}\tan\mu}{K}\frac{K^2}{\rho^2}} \\
 &&\quad \quad\leq \sqrt{2}(y^2-\tau^2)^{1/2}\dot \rho\leq \frac{1-K^2/\rho^2}{
   \left(1+\frac{K}{\rho}\right)^{2+\frac{\sqrt{3}}{\cos\mu}}
    \left(1-\frac{K}{\rho}\right)^{2-\frac{\sqrt{3}}{\cos\mu}}
    +\frac{2\sqrt{2}\tan\mu}{K}\frac{K^2}{\rho^2}}. \nn
\end{eqnarray}
Therefore, the maximal velocity of the radial null geodesics at $\rho\rightarrow K$ satisfies
\begin{eqnarray} \label{eq:d rho:R1}
&\dot \rho \propto (\rho-K), ~~ \mbox{ for } \cos\mu \geq
    \frac{\sqrt{3}}{2},\\
& \dot \rho \propto (\rho-K)^{\frac{\sqrt{3}}{\cos\mu} -1}, ~~ \mbox{ for } 0 \leq \cos \mu <\frac{\sqrt{3}}{2}. \nn
\end{eqnarray}
Integrating out the geodesic equation to calculate how much time does it take a light to escape from $\rho=K+\epsilon$ to $\rho$,
by assuming $y\gg \tau$, we get
\begin{eqnarray} \label{Delta:t:R1}
\Delta \tau \propto (\rho -K)^{2-\frac{\sqrt{3}}{\cos\mu}} ~\mbox{ for
    }  0 \leq \cos \mu <\frac{\sqrt{3}}{2},
\end{eqnarray}
and for $\cos\mu = \frac{\sqrt{3}}{2}$, we have
\begin{eqnarray} \label{Delta:t}
\Delta \tau \propto \log (\epsilon), ~~\mbox{ for } \cos\mu \geq
    \frac{\sqrt{3}}{2}.
\end{eqnarray}
This value diverges for all values of $\mu$.
Therefore, the information at the surface $\rho =K$ cannot go out of the surface through the region $y^2 < \tau^2+B^4$.
Note also that $B(\rho)$ diverges for $\cos\mu < \sqrt{3}/2$ which implies that all events at the surface $\rho=K$ satisfies the inequality  $y^2 < \tau^2+B^4$.
In this sense the surface $\rho=K$ forms an event horizon for $\cos\mu < \sqrt{3}/2$.

Now, we investigate the geodesic motion in $\mathcal{R}_2$.
For $y^2 \geq \tau^2+ B^4$, the velocity of radial null geodesics is restricted to
\begin{eqnarray} \label{dUpsilon:1}
2(y^2-\tau^2)^{1/2}\dot \Upsilon &\geq& -\left[1+\frac{B}{(y^2-\tau^2)^{1/4} }\right]^{-1}, \\
2(y^2-\tau^2)^{1/2}\dot \Upsilon &\leq& -\left[1-\frac{B}{(y^2-\tau^2)^{1/4} }\right]^{-1}, \nn
\end{eqnarray}
where the equality holds for $\dot y=0$.
Interestingly, the radial velocity is allowed to diverges for both directions.
Therefore, every information at $\rho=K$ can be transferred to the outside of the surface almost instantaneously here, which implies that the singularity at $\rho=K$ can be seen by outside observers.
This phenomenon happens because the radial coordinates $\rho$ becomes a time-like coordinates in $\mathcal{R}_2$ and the coordinate $\tau$ still remains as time-like.
In fact, this is exactly the opposite procedure of the formation of ergo-region, where time coordinate becomes spacelike and the space coordinates remain as spacelike.
In the present metric, there is no ergoregion, in which static motion is impossible, since the velocity range includes the zero radial velocity.

Finally, we summarize the geodesic motion along the light-like line $t'=z'$ at which surface the metric~(\ref{eq:met:R}) and (\ref{eq:met:F}) are ill defined.
Along the line, the velocity of radial null geodesics are calculated in Eq.~(\ref{eq:geodesic:l}). By integrating the equation, we get similar escaping time to that in Eq.~(\ref{Delta:t1}).
\end{appendix}

\begin{acknowledgments}
This work was supported in part by the Korea Research Foundation Grant funded by Korea Government (MOEHRD) (KRF-2005-075-C00009; H.-C.K.) (KRF-2006-312-C00095; J. L.), and in part by the Topical Research Program of APCTP and the National e-Science Project of KISTI.
\end{acknowledgments} \vspace{1cm}


\vspace{4cm}

\end{document}